\documentclass[showpacs,superscriptaddress,nobalancelastpage,amsmath,10pt,prl,a4paper]{revtex4} 

\usepackage{graphicx}

\begin{document}

\title{Contributions of different parts of spin-spin interactions to quantum correlations in a spin ring model in an external magnetic field}

\author{S.I. Doronin}
\email{s.i.doronin@gmail.com}
\affiliation{Institute of Problems of Chemical Physics of the Russian Academy of Sciences, 142432, Chernogolovka, Moscow Region, Russia}

\author{E.B. Fel'dman}
\affiliation{Institute of Problems of Chemical Physics of the Russian Academy of Sciences, 142432, Chernogolovka, Moscow Region, Russia} 

\author{E.I. Kuznetsova}
\affiliation{Institute of Problems of Chemical Physics of the Russian Academy of Sciences, 142432, Chernogolovka, Moscow Region, Russia}


\begin{abstract}
We study quantum correlations in a bipartite heteronuclear $(N-1)\times1$ system in an external magnetic field. The system consists of a spin ring with an arbitrary number $N-1$ of spins on the ring and one spin in its center. The spins on the ring are connected by secular dipole-dipole interactions and interact with the central spin through the Heisenberg $zz$-interaction. We show that the quantum discord, describing quantum correlations between the ring and the central spin, can be obtained analytically for this model in the high temperature approximation. The model allows us to find contributions of different parts of the spin-spin interactions to quantum correlations. We also investigate  the evolution of quantum and classical correlations at different numbers of spins.
\end{abstract}



\maketitle

\section{Introduction}

Quantum correlations in many-qubit systems are responsible for the effective work of quantum devices (in particular, quantum computers) and give them significant advantages over their classical counterparts \cite{M}. In order to create quantum devices we have to study quantum correlations and to control them. Spin systems are very suitable for the study of quantum correlations due to the developed theoretical methods \cite{G} and the experimental methods of nuclear magnetic resonance (NMR) \cite{NC,OBSFA} for the excitation, detection and coherent control of the individual spin states.

Until recently it has been accepted that entanglement is responsible for quantum correlations, and quantum devices can be created only on the basis of using materials with entangled states \cite{AFY}. However, it has turned out that quantum algorithms \cite{DSC} which significantly outperform the classical counterparts can work using mixed separable (non-entangled) states. Furthermore, it turned out that quantum non-locality can be observed in systems without entanglement \cite{BDF}. From this we can conclude that entanglement describes only a part of quantum correlations but not all of them. According to the current understanding \cite{AFY}, total (quantum and classical) correlations in a system are defined by the mutual information. The problem is how to separate the classical correlations from the quantum ones. This problem was solved independently by Henderson and Vedral \cite{HV} as well as Ollivier and Zurek \cite{OZ}. The classical correlations in a two-partite system are determined by a complete set of projective measurements carried out only over one of the subsystems \cite{HV}. Then a measure of quantum correlations (the quantum discord) is determined as the difference between the mutual information and its classical part, maximized over all possible projective measurements \cite{HV,OZ}. The quantum discord is determined completely by quantum properties of the system and equals zero for classical systems.

Computing the quantum discord is a rather tedious problem because we must optimize the quantum conditional entropy over all possible complete sets of measurements in any subsystem of the system under study \cite{HV,OZ}. It was shown \cite{YH} that computing quantum discord is NP-complete. The already developed methods [11$-$13] permit computing the discord only in two-qubit systems [14$-$17] and in the simplest three-qubit systems \cite{RS,CDF}. At the same time, it is very important to calculate the quantum discord in many-qubit systems. For example, the many-qubit quantum discord is significant for solving problems of NMR quantum information processing \cite{OBSFA}. It is well known that NMR is used to obtain an experimental implementation of quantum algorithms in few-qubit systems. The quantum entanglement vanishes at room temperatures in highly mixed states that occur in liquid-phase NMR experiments \cite{W}. This raised doubts that NMR can be used to demonstrate the advantages of quantum algorithms over their classical counterparts in multiqubit systems \cite{NC}. The quantum discord does not vanish \cite{KRMP} in mixed states in NMR experiments in contrast to the entanglement and indicates that quantum correlations exist in many-qubit systems. Experimental studies \cite{KRMP} using NMR-tomography methods \cite{OBSFA} support this conclusion. The development of methods for computing the quantum discord in many-qubit systems is therefore very relevant.

There are a lot of different NMR methods in solids which can be used for experimental investigations of quantum correlations. First of all, notice the methods of spin echo and free induction decay \cite{G,Z2}. The time evolution of signals of spin echo and free induction decay in the system of an electron (or a hetero-nucleus)  surrounded by a cloud of other nuclei can give very important information about quantum discord and its evolution \cite{Z2}. We notice also multiple quantum NMR methods \cite{BMGP} which allow us to obtain the time and temperature dependencies of quantum correlations \cite{FP}.

We have suggested \cite{CDF,DKF} a spin model for the calculation of the quantum discord. In our model, a linear chain of spins, coupled by the dipole-dipole interactions (DDI) in a strong external magnetic field, is connected with an impurity spin by the Heisenberg $zz$-interaction. This model can be used for the analysis of NMR experiments in solids, both at low and high temperatures. The developed approach \cite{CDF,DKF} allows us to study quantum correlations and their unitary evolution in the conditions of NMR experiments. Unfortunately, in that model the quantum discord can only be calculated numerically with the random mutation algorithm \cite{Ch}. The analytical calculation of the quantum discord for the model \cite{CDF} is possible only for three-qubit systems \cite{DKF}. Later we understood \cite{DFK} that it is possible to find the quantum discord analytically for many-qubit systems if one considers a spin ring instead of a spin chain used in \cite{CDF,DKF}. We note that this model can be applied not only to heteronuclear systems, but also to electron-nuclear ones. If we consider a central electron spin surrounded by a ring of nuclear spins then the model is close to the model of deterministic quantum calculations with one qubit (DQC1) \cite{DSC}.

It is necessary to emphasize that the multiqubit spin ring model can be investigated by standard tools [11$-$13] for two-part binary systems. However, we simplify those methods with an optimization of the quantum conditional entropy over the parameters of the projectors. As a result, this approach allows us to investigate contributions of different interactions to quantum correlations.

The main goal of the article is the investigation of the quantum correlations in a spin ring system and the determination of the contributions of different parts of spin-spin interactions to quantum correlations at different relation between the Larmor frequencies of the ring spins and the impurity one.

The paper is organized as follows. A spin ring system, coupled with a central spin, in the external magnetic field is presented in Sect. 2. The optimization of the quantum conditional entropy in the high temperature approximation is considered in Sect. 3. Calculations of the quantum discord are given in Sect. 4 at different relations between the Larmor frequencies of the spin subsystems. A comparison of quantum and classical correlations in the course of the system evolution is given in Sect. 5. We briefly summarize our results in Sect. 6.

\section{A bipartite many-qubit system in an external magnetic field}

We consider a ring  of nuclear spins ($I=1/2$), coupled by the DDI in a strong external magnetic field. We will call it subsystem $A$. The distances between spins of subsystem $A$ are not necessarily the same. Subsystem $A$ interacts with an impurity spin $S$, which is in the center of the ring (subsystem $B$, Fig. 1). The strong external magnetic field is perpendicular to the plane of Fig. 1.

\begin{figure}[t]
\includegraphics[width=0.4\columnwidth]{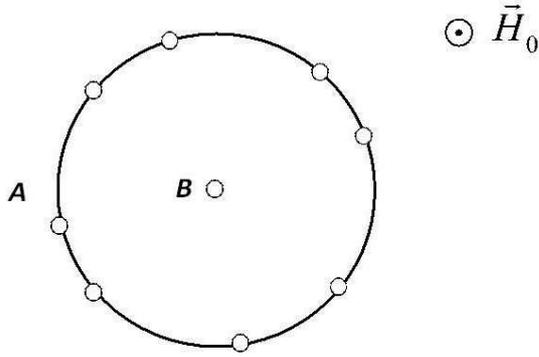}
\caption{\label{fig:F1} Many-spin systems for the investigation of the quantum discord. Subsystem $A$ is a ring of spins, coupled by the DDI. The impurity spin (in the center of the ring), connected by the $zz$-interactions with the ring spins, represents subsystem $B$.} 
\end{figure}

We will consider the problem in the high temperature approximation \cite{G}. Thes approximation is widely used in different problems of spin dynamics and magnetic resonance \cite{G}. For example, the high temperature approximation in systems of spins coupled by DDI is valid \cite{G,S} at
\begin{equation}
\beta\omega_A < 1, \quad \beta\omega_B < 1,\label{eq:11}
\end{equation}
where $\beta$ is proportional to the inverse temperature \cite{G} of the system; $\omega_A$ and $\omega_B$ are the Larmor frequencies. 

The proof is given in Ref. \cite{S} and it is based an the fact that the DDI is inversely proportional to the cube of the distances between spins. However, it is not so in the considered case. For example, $N-1$ spins interact with the central spin and all coupling constants are equal to each other. The dependence of the DDI coupling constants of the ring spins on the distances between spins is very weak and the conditions (\ref{eq:11}) are insufficiently. It is evident that the conditions should be modified. We expect that the high temperature approximation holds when
\begin{equation}
(N-1)\beta\omega_A < 1, \quad (N-1)\beta\omega_B < 1. \label{eq:12}
\end{equation}
The conditions of Eq.(\ref{eq:12}) are valid for small spin clusters. Decoherence (in particular, external noise) is determined by the relaxation times $T_1$ and $T_2$ \cite{G} which are sufficiently long for such clusters in NMR experiments \cite{VC,MACS}.
Thus, we can neglect noise in our spin systems in the calculation of quantum correlations.

In the initial moment of time, the system is in the thermodynamic equilibrium state.  In the high temperature approximation \cite{G}, the density matrix $\rho(0)$ is
\begin{equation}
\rho(0) = \frac{1}{2^N}(1 + \beta \omega_A I_z + \beta \omega_B S_z),\label{eq:1}
\end{equation}
where $z$ points in the direction of the external magnetic field; $N$ is the total number of the spins ($N-1$ is the number of the ring spins); $I_z=\sum_iI_{zi}$, and $I_{zi}$, $S_z$ are the $z$-projections of the $i$-th spin  on the ring and the impurity spin. 
Applying resonance (for spins $I$ and $S$) $90^\circ$-radiofrequency pulses of the magnetic field around the axis $y$ in the rotating reference frames (for spins $I_i, i=1,2,...,N-1$ and $S$) \cite{G}, we create conditions for the emergence of quantum correlations during the evolution of the spin system.
Formally, such initiating pulses lead to the replacement of the operators $I_{iz}$, $S_z$ with $I_{ix}$ and $S_x$, where $I_{i\alpha}$ and $S_\alpha$ ($\alpha=x,y,z$) are projections of the $i$-th ring spin and the impurity spin on the axis $\alpha$.
The subsequent evolution of the system is described by the Hamiltonian $\tilde{H}=H_{dz}+H_{zz}$.
Here the Hamiltonian $H_{dz}$ describes the secular DDI in the strong external magnetic field in the spin ring (subsystem $A$). This secular part of the DDI can be written as \cite{G}
\begin{equation}
H_{dz}=\sum_{i<j} d_{ij}\left(3I_{iz}I_{jz} - \vec{I_i}\vec{I_j}\right),\label{eq:2}
\end{equation}
where $d_{ij}$ is the DDI coupling constant of ring spins $i$, $j$; and $\vec{I_i}\vec{I_j}=I_{ix}I_{jx}+I_{iy}I_{jy}+I_{iz}I_{jz}$.
The Hamiltonian $H_{zz}$ characterizes the interaction between subsystems $A$, and $B$ and equals
\begin{equation}
H_{zz}=\sum_{i} g I_{iz}S_{z},\label{eq:3}
\end{equation}
where $g$ is the coupling constant of the ring spins with the impurity spin, which has a different gyromagnetic ratio from the ring spins. We took into account that the coupling constants of the $zz$-interactions in the considered case (see Fig. 1) are the same for all ring spins (see Eq. (\ref{eq:3})). It means that the Hamiltonians $H_{dz}$ and $H_{zz}$ commute at an arbitrary number of the ring spins. This property holds in the model with a linear spin chain instead of a spin ring only for a chain with two spins \cite{CDF}. Thus, subsystem $A$ is only subject to the dipole evolution in the considered case. Since a unitary local transformation does not change the discord, we can ignore the dipolar evolution of subsystem $A$ in our calculations. Taking into account that the evolution of the system is determined by the Heisenberg $zz$-interactions only, one can write the density matrix of the system as
\begin{equation}
\begin{array}{l}
   \rho(\tau) =e^{-iH_{zz}t}\rho(0)e^{iH_{zz}t}= \frac{1}{2^N}\Big\{1 + \beta\omega_A[I_x\cos(\tau) + 2I_yS_z\sin(\tau)] \\
					\qquad	+ \beta\omega_B[S_x\cos(2\tau I_z) + S_y\sin(2\tau I_z)]\Big\}, \label{eq:4}
      \end{array}
\end{equation}
where $\tau=gt/2$ is the dimensionless time.

The reduced density matrices $\rho_A(\tau), \rho_B(\tau)$ are necessary for the calculation of the quantum discord. They can be represented as
\begin{equation}
\rho_A(\tau) = \frac{1}{2^{N-1}}[1 + \beta\omega_A I_x\cos(\tau)], \label{eq:5}
\end{equation}
\begin{equation}
\begin{array}{l}
   \rho_B(\tau) = \frac{1}{2}\Big\{1 + \frac{1}{2^{N-1}}\beta\omega_B \Big[S_x \mathrm{Tr}[\cos(2\tau I_z)]+S_y\mathrm{Tr}[\sin(2\tau I_z)] \Big]\Big\} \\
					\qquad	= \frac{1}{2}[1 + \beta\omega_B S_x\cos^{N-1}(\tau)]. \label{eq:6}
      \end{array}
\end{equation}
We used in Eq.(\ref{eq:6}) the relationships
\begin{equation}
\mathrm{Tr}[\cos(2\tau I_z)]=2^{N-1}\cos^{N-1}(\tau), \quad \mathrm{Tr}[\sin(2\tau I_z)]=0.\label{eq:7}
\end{equation}
These relationships (\ref{eq:7}) are proved in Appendix A.

We will also use the expression for the entropy $S(\rho)$ of the system
\begin{equation}
S(\rho) = N - \frac{1}{8\ln{2}}[(N-1)\beta^2\omega^{2}_{A} + \beta^2\omega^{2}_{B}] \label{eq:8}
\end{equation}
and the expressions for the entropies $S(\rho_A)$ and $S(\rho_B)$ of subsystems $A$ and $B$ are

\begin{equation}
S(\rho_A) = N-1 - \frac{N-1}{8\ln{2}} \beta^2\omega^{2}_{A}\cos^{2}(\tau), \label{eq:9}
\end{equation}

\begin{equation}
S(\rho_B) = 1 - \frac{1}{8\ln{2}} \beta^2\omega^{2}_{B}\cos^{2(N-1)}(\tau). \label{eq:10}
\end{equation}

\section{The optimization of the quantum conditional entropy in the high temperature approximation}

According to the standard approach \cite{L,ARA} for the optimization of the quantum conditional entropy one performs a total set of projective measurements over one-qubit subsystem $B$, $\{B_k=V\Pi_kV^\dagger, k=0,1\}$, where the matrix $V \in SU(2)$ and $\Pi_k (k=0,1)$ are projectors. For our aims, it is better to write the projectors through the spin operators
\begin{equation}
\Pi_0 = \frac{1}{2} + n_xS_x+ n_yS_y+n_zS_z, \quad  \Pi_1 = \frac{1}{2} - n_xS_x- n_yS_y-n_zS_z,\label{eq:13}
\end{equation} 
where $n_x, n_y, n_z$ are the coordinates of the unit vector in the spin space ($n^{2}_{x} + n^{2}_{y} + n^{2}_{z} =1$). It is evident that $ \Pi^{2}_{k} =\Pi_k$ ($k=0,1$). Using the projectors (\ref{eq:13}) we can replace an arbitrary unitary matrix $V \in SU(2)$. At the same time, such projectors are more convenient for finding contributions of different parts of the DDI to quantum correlations.

The following relation is very useful at further calculations
\begin{equation}
\Pi_k S_\alpha \Pi_k = \frac{(-1)^k}{2} n_\alpha \Pi_k, \quad \alpha = x, y, z; \quad  k=0,1. \label{eq:14}
\end{equation} 
The density matrix $\rho(t)$ of Eq. (\ref{eq:4}) can be transformed after performing measurements with Eq. (\ref{eq:14}) as
\begin{equation}
\begin{array}{l}
\Pi_k \rho(t) \Pi_k = \frac{1}{2^N} \Big\{ I_A + \beta\omega_A[I_x\cos(\tau) + (-1)^k n_z I_y\sin(\tau)] \\
					\qquad	+ (-1)^k\frac{1}{2}\beta\omega_B[n_x\cos(2\tau I_z) + n_y\sin(2\tau I_z)]\Big\}\otimes \Pi_k, \quad  k=0,1, \label{eq:15}
\end{array}					
\end{equation}
where $I_A$ is a unit operator in the spin space of subsystem $A$. Thus, one can find that the whole system is described by the ensemble of the states $\{p_k, \rho_k \} (k=0,1)$ after the measurements, where
\begin{equation}
\begin{array}{l}
p_0= \mathrm{Tr}\{\Pi_0\rho(t)\Pi_0\}=\frac{1}{2}+\frac{1}{4}n_x\beta\omega_B\cos^{N-1}(\tau), \\
p_1= \mathrm{Tr}\{\Pi_1\rho(t)\Pi_1\}=\frac{1}{2}-\frac{1}{4}n_x\beta\omega_B\cos^{N-1}(\tau),\label{eq:16}
      \end{array}
\end{equation}
and the matrices $\rho_0, \rho_1 $ are
\begin{equation}
\begin{array}{l}
  \rho_0 = \frac{1}{2^N p_0}\Big\{1 + \beta\omega_A [I_x \cos(\tau)+ n_zI_y\sin(\tau)] \\
	\qquad +\frac{1}{2}\beta\omega_B[n_x\cos(2\tau I_z) + n_y\sin(2\tau I_z)]\Big\} , \\
	\rho_1 = \frac{1}{2^N p_1}\Big\{1 + \beta\omega_A [I_x \cos(\tau)- n_zI_y\sin(\tau)] \\
	\qquad -\frac{1}{2}\beta\omega_B[n_x\cos(2\tau I_z) + n_y\sin(2\tau I_z)]\Big\}.
	\label{eq:17}
      \end{array}
\end{equation}
The results of Appendix A are used again at calculations of $p_0$ and $p_1$ in Eqs. (\ref{eq:16}). Then the conditional quantum entropy $S_{cond}$ after the measurements over subsystem $B$ can be written \cite{L,ARA} as
\begin{equation}
S_{cond}= p_0S(\rho_0)+p_1S(\rho_1),\label{eq:18}
\end{equation}
where $S(\rho_k)=-\mathrm{Tr}[\rho_k\log_2\rho_k]$.
It is convenient to introduce the parameters $u$ and $v$, describing the high temperature approximation (see Eq. (\ref{eq:12}))
\begin{equation}
u=\frac{1}{2}(N-1)\beta\omega_B < 1, \quad v=\frac{1}{2}(N-1)\beta\omega_A < 1.\label{eq:19}
\end{equation}
The calculation of the quantum conditional entropy (\ref{eq:18}) with equations (\ref{eq:16}),(\ref{eq:17}), (\ref{eq:19}) up to the terms of the orders $u^2$ and $v^2$  leads to the formula
\begin{equation}
\begin{array}{l}
   S_{cond} = -\frac{1}{2\ln2(N-1)^2}\Big\{n^{2}_{x} \Big[u^2\frac{1+\cos^{N-1}(2\tau)-2\cos^{2(N-1)}(\tau)}{2} - (N-1)v^2\sin^2(\tau)\Big] \\
\qquad	+ n^{2}_{y} \Big[u^2\frac{1-\cos^{N-1}(2\tau)}{2} - (N-1)v^2\sin^2(\tau)\Big] + a(u,v)\Big\}, \label{eq:20}
      \end{array}
\end{equation}
where the function $a(u,v)=(N-1)v^2-2(N-1)^3\ln2$ does not depend on $n_\alpha (\alpha=x,y,z)$.

\section{The quantum discord at different relations between the Larmor frequencies of the spin subsystems}

In this section we demonstrate that quantum correlations depend on relations between the Larmor frequencies of the ring spins and the central one. First, we consider the case of $v>u$, i.e., the Larmor frequency of the ring spins exceeds those of the central spin. We show in Appendix B that the coefficients of equation (\ref{eq:20}) at $n_x$ and $n_y$ are positive in this case. It means that the quantum conditional entropy of equation (\ref{eq:18}) achieves the minimal value at
\begin{equation}
n_x=n_y=0, \quad |n_z|=1.\label{eq:21}
\end{equation}
In this case (see Eq. (\ref{eq:15})) only $I_yS_z$-interactions can give a contribution to quantum correlations due to the term which is proportional to $n_zI_y$.

Using Eqs.(\ref{eq:8}), (\ref{eq:10}), (\ref{eq:20}) and (\ref{eq:21}), one can obtain an expression for the quantum discord $D$ \cite{HV,OZ}, which is the entropic measure of the quantum correlations
\begin{equation}
D=\frac{u^2}{2\ln2(N-1)^2} \Big[ 1 - \cos^{2(N-1)}(\tau)\Big] .\label{eq:22}
\end{equation}
Naturally, the quantum discord here is determined by the lower Larmor frequency $u$ of the impurity spin. We notice also that formula (\ref{eq:22}) is valid for such numbers of the ring spins when the conditions (\ref{eq:12}) of the high temperature approximation \cite{G,S} hold.

The dependence of the quantum discord on the dimensionless evolution time $\tau$ at the different numbers of the spins with $u/(N-1)=\frac{1}{2}\beta\omega_B=0.015$ is shown in figure 2.

\begin{figure}[t]
\includegraphics[width=0.5\columnwidth]{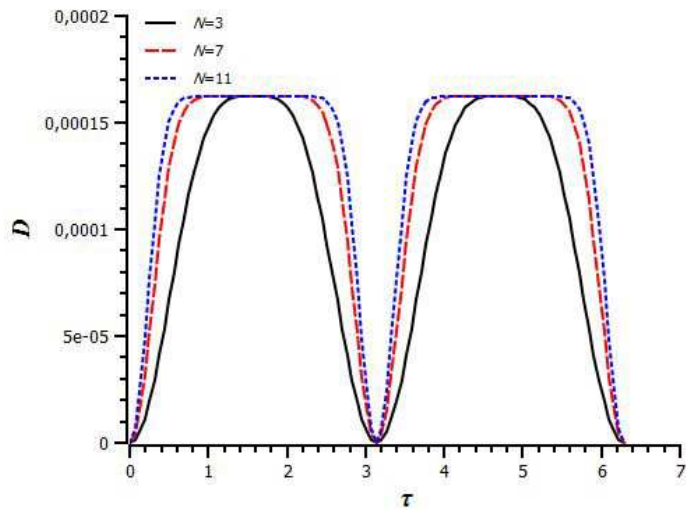}
\caption{\label{fig:F2} The dependence of the quantum discord on the evolution time at different numbers of the  spins ($N=3, 7, 11$) with $u=0.015$. The minimal value of the conditional entropy is realized at $n_x=n_y=0$ and $|n_z|=1$.} 
\end{figure}

The case $u>v$ is very difficult for an analytical investigation. However, it is possible to perform the optimization of the quantum conditional entropy at some additional conditions. Let
\begin{equation}
u^2>(N-1)v^2, \quad 0<\tau<\pi/4, \quad N-1\geq2. \label{eq:23}
\end{equation}
Then we prove in Appendix C that the quantum conditional entropy of equation (\ref{eq:18}) achieves the minimal value at
\begin{equation}
n_x=n_z=0, \quad |n_y|=1.\label{eq:24}
\end{equation}
We can conclude again from equation (\ref{eq:15}) that $I_zS_y$-interactions are responsible for quantum correlations at the conditions (\ref{eq:23}). Using Eqs.(\ref{eq:8}), (\ref{eq:10}), (\ref{eq:20}) and (\ref{eq:24}), one can calculate the corresponding quantum discord:
\begin{equation}
\begin{array}{l}
D=\frac{1}{2\ln2(N-1)^2} \Big\{ (N-1)v^2\sin^2(\tau)+\frac{u^2}{2}\Big[\cos^{N-1}(2\tau)+1-2\cos^{2(N-1)}(\tau)\Big]\Big\}. \label{eq:25}
      \end{array}
\end{equation}
It is worth to notice that $I_zS_y$- interactions are also responsible for quantum correlations not only at the conditions (\ref{eq:23}) but also when
\begin{equation}
u^2>\frac{4}{3}\frac{N-1}{1-3^{-(N-1)}}v^2, \quad \pi/4\leq\tau<\arctan(\sqrt{2}), \label{eq:_26}
\end{equation}
Then again the quantum conditional entropy achieves its minimal value when Eq. (\ref{eq:24}) holds, and the quantum discord is determined by formula (\ref{eq:25}).

Finally, we show in Appendix C that at the conditions 
\begin{equation}
u^2>2(N-1)v^2, \quad N\geq 3 \,\, \textrm{is odd}, \quad \arctan(\sqrt{2})<\tau<\pi/2, \label{eq:26}
\end{equation}
the quantum conditional entropy of equation (\ref{eq:18}) achieves its minimal value at
\begin{equation}
n_y=n_z=0, \quad |n_x|=1.\label{eq:27}
\end{equation}
We find with equation (\ref{eq:15}) that $I_zS_x$-interactions are responsible for quantum correlations in this case and the quantum discord is
\begin{equation}
\begin{array}{l}
D=\frac{1}{2\ln2(N-1)^2} \Big\{ (N-1)v^2\sin^2(\tau)+\frac{u^2}{2}\Big[1-\cos^{N-1}(2\tau)\Big]\Big\}.\label{eq:28}
      \end{array}
\end{equation}

\begin{figure}[t]
\includegraphics[width=0.5\columnwidth]{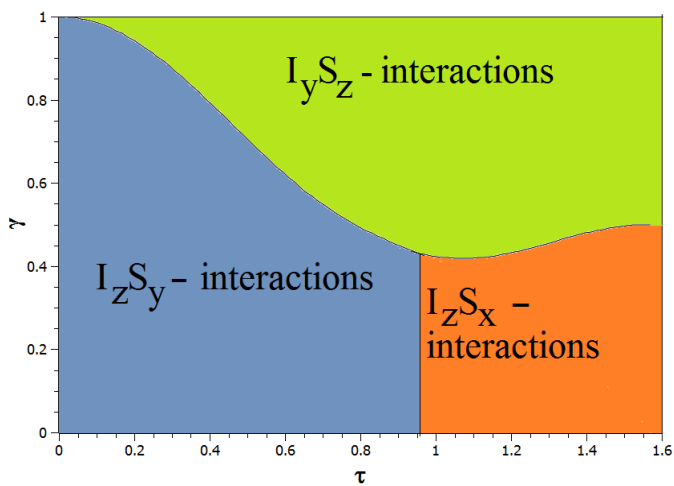}
\caption{\label{fig:F3} The regions of the parameters $\gamma=\omega_A/\omega_B$ and the dimensionless evolution time $\tau$, where $I_yS_z$-, $I_zS_y$-, $I_zS_x$- spin-spin interactions are responsible for quantum correlations at $N=5$.}
\end{figure}

In Fig. 3 we demonstrate the regions, where $I_yS_z$-, $I_zS_y$- and $I_zS_x$-interactions are responsible for quantum correlations at different $\gamma=\omega_A/\omega_B$ and the evolution time $\tau$ for $N=5$.

\section{A comparison of quantum and classical correlations in the course of spin evolution}

Classical correlations can be obtained from the maximal information about subsystem $A$ after measurements on subsystem $B$ \cite{HV,AFY}. The total correlations $I(\rho)$ are determined by
\begin{equation}
I(\rho)=S(\rho_A)+S(\rho_B)-S(\rho), \label{eq:29}
\end{equation}
where the entropies $S(\rho_A)$, $S(\rho_B)$, $S(\rho)$ are determined by the expressions (\ref{eq:8}), (\ref{eq:9}), (\ref{eq:10}). Taking into account formulas (\ref{eq:22}), (\ref{eq:25}), (\ref{eq:28}) for the quantum discord, one can obtain the expressions for classical correlations $C$ at the different conditions for the minimal value of the quantum conditional entropy:
\begin{equation}
C=\frac{1}{2\ln2(N-1)}v^2\sin^2(\tau), \quad \textrm{for} \quad |n_z|=1, \quad n_x=n_y=0; \label{eq:30}
\end{equation}
\begin{equation}
\begin{array}{l}
C=\frac{u^2}{4\ln2(N-1)^2} \Big[1+\cos^{N-1}(2\tau)-2\cos^{2(N-1)}(\tau)\Big], \\
 \quad \textrm{for} \quad |n_x|=1, \quad n_y=n_z=0; \label{eq:31}
      \end{array}
\end{equation}
\begin{equation}
C=\frac{u^2}{4\ln2(N-1)^2} \Big[1-\cos^{N-1}(2\tau)\Big], \quad \textrm{for} \quad |n_y|=1, \quad n_x=n_z=0. \label{eq:32}
\end{equation}

\begin{figure}[t]
\includegraphics[width=1.0\columnwidth]{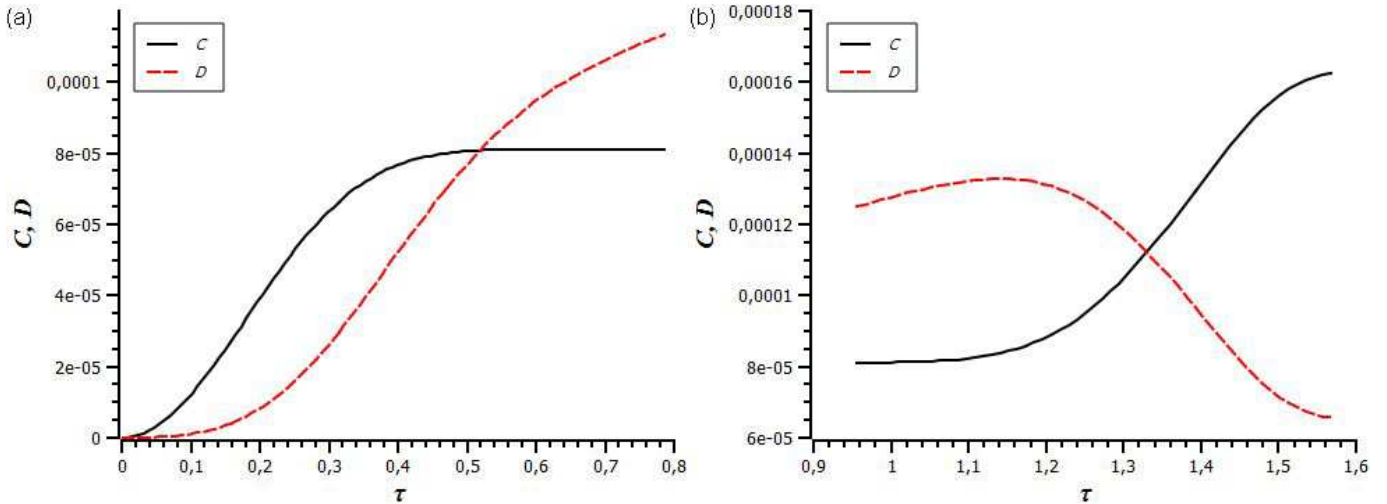}
\caption{\label{fig:F4} A comparison of quantum and classical correlations in the course of the evolution of the spin system at $N=9$;
a) an initial period of the evolution $0<\tau<\pi/4$;
b) the evolution on the time interval $\arctan{(\sqrt{2})}<\tau<\pi/2$.}
\end{figure}

We can compare quantum and classical correlations in the course of the evolution at different numbers of spins. Such comparison is presented for the system, consisting of $N=9$ spins, in Fig. 4. Fig. 4a demonstrates that classical correlations exceed quantum ones at the initial period of the evolution. However, the quantum correlations increase with time and exceed classical at $0.521<\tau<\pi/4$. The evolution of quantum and classical correlations on the time interval $[\arctan{(\sqrt{2})}, \, \pi/2]$ is given in Fig.4b. Quantum correlations exceed classical ones on the main part of this interval. However, quantum correlations rapidly decrease and classical correlations increase at the end of the interval $[\arctan{(\sqrt{2})}, \, \pi/2]$.

\section{Conclusions}

Our work is aimed at the analytical computation of the quantum discord in a system of interacting spins in a strong external magnetic field. The spin dynamics of this system can be investigated experimentally by NMR methods.

Our results demonstrate that there is a non-zero quantum discord at room temperature in the considered spin system. It is relevant for NMR in liquids where entanglement is absent at room temperature. Our results mean that there are quantum correlations in the considered system at room temperature and possibilities of NMR methods for creating quantum devices and performing quantum calculations are not exhausted.

We modified the standard methods \cite{L,ARA} for the calculation of the quantum discord in bipartite systems. Our idea is to use projectors for quantum measurements instead of arbitrary unitary transformations in order to optimize the quantum conditional entropy over parameters of those projectors.

We showed that different parts of the spin-spin interactions are responsible for quantum correlations at different relations between the Larmor frequencies of the ring and impurity spins. It is very important that the model considered in this paper allows us to calculate the quantum discord analytically and we obtained the analytical formulas for the quantum discord at arbitrary relations between the Larmor frequencies of the ring spins and the impurity one. We have also investigated the time evolution of the quantum discord in the spin ring system with a central spin in the external magnetic field. The quantum and classical correlations were compared in the course of the spin system evolution.

Our model is also likely to be useful in the investigation of quantum and classical correlations at low temperatures where the quantum discord is significantly larger than at room temperature.

\begin{acknowledgements}
The work is supported by the Russian Foundation for Basic Research (Grants No. 13-03-00017, No. 13-03-12418 and No. 15-07-07928) and the Program of the Presidium of RAS No. 32 ``Electron spin resonance, spin-dependent electron effects and spin technologies''.
\end{acknowledgements} 

\appendix
\setcounter{section}{1}
\section*{Appendix A}

The ring contains $N-1$ spins and one can write that

\begin{equation}
\mathrm{Tr}\{\cos (2\tau I_z)\}=\sum_{k=0}^{N-1}C_{N-1}^k\cos \left[2\tau \left(\frac{N-1}{2}-k\right)\right],
\end{equation}
where $C_{N-1}^k={N-1 \choose k}=\frac{(N-1)!}{k!(N-1-k)!}$. It is suitable to use the complex values:
\begin{equation}
\begin{array}{l}
\sum_{k=0}^{N-1}C_{N-1}^k\cos \left[2\tau \left(\frac{N-1}{2}-k\right)\right]=\mathrm{Re}\left\{\sum_{k=0}^{N-1}C_{N-1}^k e^{i\left[2\tau \left(\frac{N-1}{2}-k\right)\right]}  \right\} \\
=\mathrm{Re}\left\{ e^{i\tau (N-1)} \sum_{k=0}^{N-1} C_{N-1}^k e^{-2i\tau k}\right\}
\end{array}
\end{equation}
Applying the binomial theorem one obtains
\begin{equation}
\begin{array}{l}
\mathrm{Re}\left\{ e^{i\tau (N-1)} \sum_{k=0}^{N-1} C_{N-1}^k e^{-2i\tau k}\right\}=\mathrm{Re}\left\{ e^{i\tau(N-1)}(1+e^{-2i\tau})^{N-1}  \right\} \\
=\mathrm{Re}\left\{ (e^{i\tau}+e^{-i\tau})^{N-1}  \right\}=2^{N-1}\cos^{N-1}(\tau).
\end{array}
\end{equation}
Now consider $\mathrm{Tr}\{\sin (2\tau I_z)\}$ and perform the rotation $ e^{i\pi I_x}$ of the operator $\sin (2\tau I_z)$:
\begin{equation}
\begin{array}{l}
\mathrm{Tr}\{\sin (2\tau I_z)\}=\mathrm{Tr}\{e^{i\pi I_x}\sin (2\tau I_z)e^{-i\pi I_x}\}\\
=-\mathrm{Tr}\{\sin (2\tau I_z)\}.
\end{array}
\end{equation}
It means that $\mathrm{Tr}\{\sin (2\tau I_z)\}=0$. The relations (\ref{eq:7}) are proved.

\appendix
\setcounter{section}{2}
\section*{Appendix B}

We need to prove (see Eq.(\ref{eq:20})) that
\begin{equation} \label{uslovie1}
1+\cos^{N-1}(2\tau)-2\cos^{2(N-1)}(\tau)\leq 2(N-1)\sin^2 (\tau).
\end{equation}
The left hand side of  Eq.(\ref{uslovie1}) can be rewritten as
\begin{equation}
\begin{array}{l} 
1+\cos^{N-1}(2\tau)-2\cos^{2(N-1)}(\tau) \\ 
=1-\cos^{2(N-1)}(\tau)+\cos^{N-1}(2\tau)-\cos^{2(N-1)}(\tau) \\
=\sin^2(\tau)\left(1+\cos^2(\tau)+\cos^4(\tau)+...+\cos^{2(N-2)}(\tau)\right) \\
+\Big(\cos(2\tau)-\cos^{2}(\tau)\Big)\Big(\cos^{N-2}(2\tau)+\cos^{N-3}(2\tau)\cos^2(\tau) \\ 
+...+\cos^{2(N-2)}(\tau)\Big) \leq 
\sin^2(\tau)(N-1)+\sin^2(\tau)(N-1)\label{stroka}
\end{array}
\end{equation}
The statement follows from the inequality (\ref{stroka}).

Further we need to prove (see Eq. \ref{eq:20}) that 
\begin{equation}\label{uslovie2}
1-\cos^{N-1}(2\tau)\leq 2(N-1)\sin^2 (\tau). \end{equation}
We can transform the left hand side of Eq. (\ref{uslovie2}) in the following way
\begin{equation}
\begin{array}{l} 
1-\cos^{N-1}(2\tau)=\Big(1-\cos(2\tau)\Big)\Big(1+\cos(2\tau)+\cos^2(2\tau) \\
+... + \cos^{N-2}(2\tau)\Big)\leq 2\sin^2 (\tau) (N-1)\label{stroka2} 
\end{array}
\end{equation}
The statement (\ref{uslovie2}) follows from the inequality (\ref{stroka2}). The  inequalities (\ref{uslovie1}) and (\ref{uslovie2}) lead to the conditions (\ref{eq:21}) for the minimal value of the conditional entropy of Eq. (\ref{eq:20}).

\appendix
\setcounter{section}{3}
\section*{Appendix C}

We will show that the coefficient of $n_y^2$ in the curly brackets of Eq. (\ref{eq:20}) exceeds the one of $n_x^2$ at the conditions (\ref{eq:23}). One obtains from Eq. (\ref{eq:20}) that we need to prove that
\begin{equation}
\begin{array}{l} 
u^2\frac{1-\cos^{N-1}(2\tau)}{2}-(N-1)v^2\sin^2 (\tau) > u^2\frac{1+\cos^{N-1}(2\tau)-2\cos^{2(N-1)}(\tau)}{2}  \\ 
-(N-1)v^2\sin^2 (\tau). \label{eq:C1}
\end{array}
\end{equation}
It is evident that Eq. (\ref{eq:C1}) can be transformed to
\begin{equation}
1 > [1-\tan^2 (\tau)]^{N-1}. \label{eq:C2}
\end{equation}
Eq. (\ref{eq:C2}) is valid at the conditions (\ref{eq:23}).
Now we prove that the coefficient of $n_y^2$ in the curly brackets of Eq. (\ref{eq:20}) is non-negative at the conditions (\ref{eq:23}). In fact, we should prove that
\begin{equation}
\frac{1-\cos^{N-1}(2\tau)}{2} > \sin^2 (\tau). \label{eq:C3}
\end{equation}
Eq. (\ref{eq:C3}) is equivalent to
\begin{equation}
\cos(2\tau) > \cos^{N-1}(2\tau) \label{eq:C4}
\end{equation}
and Eq. (\ref{eq:C4}) holds due to the conditions (\ref{eq:23}). One can conclude from Eq. (\ref{eq:20}) that the minimal value of the quantum conditional entropy is achieved at the parameters (\ref{eq:24}).

Next, we will show that the coefficient of $n_x^2$ in the figure bracket of Eq. (\ref{eq:20}) exceeds the one of $n_y^2$ at the conditions (\ref{eq:26}).
In this case we can obtain  that this statement is valid, if
\begin{equation}
[1-\tan^2 (\tau)]^2 > 1. \label{eq:C5}
\end{equation}
However, it is really valid at the conditions (\ref{eq:26}). We show now that the coefficient of $n_x^2$ in the curly bracket of (\ref{eq:20}) is non-negative. We would like to prove that
\begin{equation}
u^2\frac{1+\cos^{N-1}(2\tau)-2\cos^{2(N-1)}(2\tau)}{2} - (N-1)v^2\sin^2 (\tau) > 0. \label{eq:C6}
\end{equation}
Using the conditions (\ref{eq:26}), one can obtain from Eq. (\ref{eq:C6}) that
\begin{equation}
[\cos^2(\tau)-\cos^{(N-1)}(\tau)] + [\cos^{N-1}(2\tau)-\cos^{2(N-1)}(\tau)] > 0. \label{eq:C7}
\end{equation}
The expression in the first square bracket is non-negative at $N \geq 3$. The second square bracket can be transformed as
\begin{equation}
\cos^{N-1}(2\tau)-\cos^{2(N-1)}(\tau)=\cos^{2(N-1)}(\tau)\Big[\Big(1-\tan^2(\tau)\Big)^{N-1} - 1\Big] > 0. \label{eq:C8}
\end{equation}
This term is positive because $N-1$ is even and $\arctan{(\sqrt{2})}<\tau<\pi/2$ due to the conditions (\ref{eq:26}). In fact, we proved that the quantum conditional entropy of (\ref{eq:18}) achieves its minimal value at the conditions (\ref{eq:27}).

\end{document}